\documentclass[superscriptaddress]{revtex4}
\usepackage{amssymb}
\usepackage{amsmath}
\usepackage{graphicx}

\begin{document}
\def\be{\begin{eqnarray}}
\def\ee{\end{eqnarray}}

\title{Entanglement-breaking of quantum dynamical channels}
\author{Long-Mei Yang}
\affiliation{School of Mathematical Sciences,  Capital Normal University,  Beijing 100048,  China}
\author{Tao Li}
\thanks{Corresponding author: litao@btbu.edu.cn}
\affiliation{School of Science, Beijing Technology and Business University, Beijing 100048, China}
\author{Shao-Ming Fei}
\thanks{Corresponding author: feishm@cnu.edu.cn}
\affiliation{School of Mathematical Sciences,  Capital Normal University,  Beijing 100048,  China}
\affiliation{Max Planck Institute for Mathematics in the Sciences, Leipzig 04103, Germany}
\author{Zhi-Xi Wang}
\thanks{Corresponding author: wangzhx@cnu.edu.cn}
\affiliation{School of Mathematical Sciences,  Capital Normal University,  Beijing 100048,  China}

\begin{abstract}
Entanglement is a key issue in the quantum physics which gives rise to resources for achieving tasks that are not possible within the realm of classical physics. Quantum entanglement varies with the evolution of the quantum systems. It is of significance to investigate the entanglement dynamics in terms of quantum channels. We study the entanglement-breaking channels and present the necessary and sufficient conditions for a quantum channel to an entanglement-breaking one for qubit systems. Furthermore, a concept of strong entanglement-breaking channel is introduced. The amendment of entanglement-breaking channels is also studied.
\end{abstract}

\maketitle

\section{Introduction}
Originated from the superposition principle, entanglement was first recognized as a ¡°spooky¡± feature of quantum mechanics by Einstein, Podolsky, Rosen and Schr$\ddot{\mathrm{o}}$dinger. This feature implies the existence of global states which cannot be written as a product of the states of individual subsystems. Attributed to the development of quantum information science over the last decades, many quantum information protocols \cite{quantum} have been proposed, in which the entangled states are used as important resources that may be exploited to achieve tasks that are not possible within the realm of classical physics.

Any physical processes correspond to completely
positive and trace-preserving (CPTP) maps called quantum channels.
Quantum channels may disturb the transmitted messages by gradually degrading the information along with their structures \cite{trans,send}. It is of considerable interest to study entanglement evolution under noisy channels. Entanglement-breaking (EB) channels \cite{M.H,qubit,cost}, one type of noisy channels, lie the heart of such tasks. The entanglement-breaking channels are so noisy that any output states are always separable, which are useless for entanglement distributions, even exploiting distillation techniques \cite{purification}.

Although entanglement-breaking channels have drawn much attention, there are still no rigorous and operational approaches to determine whether a channel is EB or not. We investigate the entanglement-breaking channels and answer such question for qubit systems.
Following a natural question that how strength an entanglement-breaking channel can be or can it be amended, we propose a concept -- strong entanglement-breaking channels.

An entanglement-breaking channel $\Phi$ might be amended through $\tilde{\Phi}\circ\Phi_U\circ\tilde{\Phi}\circ\cdots\circ\Phi_U\circ\tilde{\Phi}$,
with $\Phi_U$ a proper unitary operation, and $\Phi$  consecutive applications of a given map $\tilde{\Phi}$ repeated $n$ times, $\Phi=\tilde{\Phi}^n=\underbrace{\tilde{\Phi}\circ\tilde{\Phi}\circ\cdots\circ\tilde{\Phi}}\limits_{n \ \mathrm{times}}$ \cite{Amending}. While we find that not all the entanglement-breaking channels can be amended by such method, we present a theorem to show that there indeed exist strong entanglement-breaking channels.

This paper is organized as follows.
In Section $2$ we introduce some preliminary notions.
In Sections $3$ we investigate unital and non-unital entanglement-breaking channels, respectively. And
some necessary and sufficient conditions for an entanglement-breaking channel are presented.
We analyze entanglement-breaking of Markovian semigroup dynamics in Section 4.
In Section 5, we propose the concept of strong entanglement-breaking and  give an example to discuss its amendment via global operations.
We conclude in section $6$.

\section{Preliminaries}
We now introduce some relevant concepts. A quantum channel $\Phi$, completely positive and trace-preserving (CPTP) map, maps a quantum state $\rho$ in Hilbert space $\mathcal{H}$ to $\Phi(\rho)=\sum\limits_nK_n\rho K_n^{\dag}$, where $K_n$ are operators on $\mathcal{H}$ satisfying $\sum\limits_nK_n^{\dag}K_n=I$, with $I$ the identity operator.  Recall that the identity and Pauli matrices form a basis for $\mathcal{H}$ so that any qubit state $\rho$ can be written as $\displaystyle\frac{1}{2}(I+\roarrow{\textbf{\textit{r}}}\cdot\mathbf{\sigma})$, where $\sigma=(\sigma_x,\sigma_y,\sigma_z)$ denotes the vector of Pauli matrices and  $\roarrow{\textbf{\textit{r}}}\in\mathbb{R}^3$, $|\roarrow{\textbf{\textit{r}}}|\leqslant1$. Then, the channel $\Phi$ can be characterized by a unique $4\times 4$ matrix
$\left(
 \begin{array}{cc}
  1 & \roarrow{\mathbf{0}} \\
  \roarrow{\textbf{\textit{n}}} & M\\
  \end{array}
  \right)$,
where $M$ is a $ 3\times3$ matrix ($\roarrow{\mathbf{0}}$ and $\roarrow{\textbf{\textit{n}}}$ are three dimensional row and column vectors, respectively) such that
$$
\Phi[\frac{1}{2}(I+\roarrow{\textbf{\textit{r}}}\cdot\mathbf{\sigma})]=\frac{1}{2}I
+(\roarrow{\textbf{\textit{n}}} +M\frac{1}{2}\roarrow{\textbf{\textit{r}}})\cdot\sigma.
$$

Furthermore, King and Ruskai \cite{C.King} showed that up to a unitary equivalence the original 12-parametric set of single-qubit channels can be reduced to a 6-parametric set of channels of the form
\begin{equation}\label{eqphi}
\left(
  \begin{array}{cccc}
    1 & 0 & 0 & 0 \\
    n_1 & \lambda_1 & 0 & 0 \\
    n_2 & 0 & \lambda_2 & 0 \\
    n_3 & 0 & 0 & \lambda_3 \\
  \end{array}
\right),
\end{equation}
where $|\lambda_k|\leqslant1$ are the singular values of $M$. It can be seen that a channel is unital if and only if $\roarrow{\textbf{\textit{n}}}=\roarrow{\textbf{0}}$.

Moreover, $\Phi$ is an entanglement-breaking (EB) channel \cite{M.H} if the state $\Phi\otimes I(\rho)$ is separable for all $\rho$. It has been proved by Horodecki {\it et al.} that $\Phi$ is entanglement-breaking if and only if $\Phi\otimes I$ maps a maximally entangled state to a separable one. Without loss of generality, in the following we consider the maximally entangled state,
$$
|\psi\rangle\langle\psi|=\displaystyle\frac{1}{4}(I\otimes I-\sigma_x\otimes\sigma_x-\sigma_y\otimes\sigma_y-\sigma_z\otimes\sigma_z),
$$
to study the entanglement-breaking channels.

\section{Unital and non-unital channels}
We first study the necessary and sufficient conditions for unital channels $\Phi$ to be
EB ones. Under the channels $\Phi\otimes I$, the state $|\psi\rangle\langle\psi|$ becomes $\tilde{\rho}$:
\small
\begin{align}
\text{\normalsize $\tilde{\rho}$}&=\text{\normalsize $\Phi\otimes I(|\psi\rangle\langle\psi|)$}\nonumber\\[2mm]
&=\text{\normalsize $\frac{1}{4}$}\left(
              \begin{array}{cccc}
                1-\lambda_3 & 0 & 0 & -\lambda_1+\lambda_2 \\
                0 & 1+\lambda_3 & -\lambda_1-\lambda_2 & 0 \\
                0 & -\lambda_1-\lambda_2 & 1+\lambda_3 & 0 \\
                -\lambda_1+\lambda_2 & 0 & 0 & 1-\lambda_3 \\
              \end{array}
            \right).
\end{align}
\normalsize
The eigenvalues of $\tilde{\rho}$ are
\begin{equation}\label{egv1}
\begin{array}{rl}
\displaystyle \frac{1}{4}(1+\lambda_1-\lambda_2-\lambda_3), ~~\displaystyle  \frac{1}{4}(1-\lambda_1+\lambda_2-\lambda_3), \\ [3mm]
\displaystyle \frac{1}{4}(1-\lambda_1-\lambda_2+\lambda_3), ~~\displaystyle  \frac{1}{4}(1+\lambda_1+\lambda_2+\lambda_3),
\end{array}
\end{equation}
and the eigenvalues of the partially transposed matrix $\tilde{\rho}^{T_2}=(I\otimes T)\tilde{\rho}$, where $T$ denotes transpose, are given by
\begin{equation}\label{egv2}
\begin{array}{rl}
\displaystyle\frac{1}{4}(1-\lambda_1-\lambda_2-\lambda_3), ~~\displaystyle \frac{1}{4}(1-\lambda_1+\lambda_2-\lambda_3), \\ [3mm]
\displaystyle\frac{1}{4}(1+\lambda_1-\lambda_2+\lambda_3), ~~\displaystyle \frac{1}{4}(1+\lambda_1+\lambda_2+\lambda_3).
\end{array}
\end{equation}
It is easy to see that all eigenvalues of $\tilde{\rho}$ and $\tilde{\rho}^{T_2}$, given by \eqref{egv1} and \eqref{egv2}, are nonnegative iff
\small
\begin{equation}\label{rho}
\min\{(1-\lambda_3)^2, (1+\lambda_3)^2\}\geqslant\max\{(\lambda_1-\lambda_2)^2, (\lambda_1+\lambda_2)^2\},
\end{equation}
\normalsize
which is equivalent to
\begin{equation}\label{rhoT2}
|\lambda_1|+|\lambda_2|+|\lambda_3|\leqslant1.
\end{equation}
Here if one or more $\lambda_i=0$, then $\rho$ and $\tilde{\rho}$ have the same eigenvalues.

As $\Phi$ is an EB channel iff $\tilde{\rho}^{T_2}$ is a positive operator, according to the positive partial transposition criteria, we have the following theorem.

\noindent $\textbf{Theorem 1.}$
{\rm (i)} If $\Phi$ given by \eqref{eqphi} is a unital channel with at least one vanishing $\lambda_i$, then $\Phi$ is an entanglement breaking channel.\\
{\rm (ii)} A unital channel $\Phi$ defined by \eqref{eqphi} is an entanglement-breaking channel iff either \eqref{rho} or \eqref{rhoT2} is satisfied.

Next, we consider the non-unital channels. Let $\Phi$ be a non-unital channel. Accordingly, we have
\begin{align*}
\tilde{\rho}&=\Phi\otimes I(|\psi\rangle\langle\psi|) \nonumber\\
&=\frac{1}{4}\left(\footnotesize
              \begin{matrix}
                1+n_1-\lambda_3 & 0 & n_1-in_2 & -\lambda_1+\lambda_2 \\
                0 & 1+n_3+\lambda_3 & -\lambda_1-\lambda_2 & n_1-in_2 \\
                n_1+in_y & -\lambda_1-\lambda_2 & 1-n_3+\lambda_3 & 0 \\
                -\lambda_1+\lambda_2 & n_1+in_2 & 0 & 1-n_3-\lambda_3
              \end{matrix}
            \right).
\end{align*}
\normalsize
In general it is hard to compute the eigenvalues of $\tilde{\rho}$. However, if $\lambda_1=0$, the eigenvalues of $\tilde{\rho}$ and $\tilde{\rho}^{T_2}$ are the same:
\begin{equation}\label{rho1}
\begin{array}{rl}
\displaystyle
\frac{1}{4}(1-\sqrt{|\roarrow{\boldsymbol{\lambda}}|^2+|\roarrow{\textbf{\textit{n}}}|^2-2\sqrt{\lambda_2^2\lambda_3^2+\lambda_2^2n_2^2+\lambda_3^2n_3^2}}), \\ [3mm]
\displaystyle
\frac{1}{4}(1+\sqrt{|\roarrow{\boldsymbol{\lambda}}|^2+|\roarrow{\textbf{\textit{n}}}|^2-2\sqrt{\lambda_2^2\lambda_3^2+\lambda_2^2n_2^2+\lambda_3^2n_3^2}}),\\[3mm]
\displaystyle
\frac{1}{4}(1-\sqrt{|\roarrow{\boldsymbol{\lambda}}|^2+|\roarrow{\textbf{\textit{n}}}|^2+2\sqrt{\lambda_2^2\lambda_3^2+\lambda_2^2n_2^2+\lambda_3^2n_3^2}}), \\ [3mm]
\displaystyle \frac{1}{4}(1+\sqrt{|\roarrow{\boldsymbol{\lambda}}|^2+|\roarrow{\textbf{\textit{n}}}|^2+2\sqrt{\lambda_2^2\lambda_3^2+\lambda_2^2n_2^2+\lambda_3^2n_3^2}}),
\end{array}
\end{equation}
where $|\roarrow{\boldsymbol{\lambda}}|=\sqrt{\lambda_1^2+\lambda_2^2+\lambda_3^2}$ and $|\roarrow{\textbf{\textit{n}}}|=\sqrt{n_1^2+n_2^2+n_3^2}$. Therefore, $\Phi$ is an EB channel.
Analogously, one can show that $\Phi$ is also an EB channel for the cases $\lambda_2=0$ and $\lambda_3=0$.

Similar to Theorem 1 for unital channel case, we have

\noindent $\textbf{Theorem 2.}$  If $\Phi$ is a non-unital channel defined by \eqref{eqphi} such that at least one $\lambda_k=0$ for $k=1,2,3$, then $\Phi$ is an EB channel.

\noindent $\textbf{Theorem 3.}$
\label{EB2} If $n_{i_1}=n_{i_2}=0$ for $i_1,\ i_2,\ i_3=\{1,\ 2,\ 3\}$,
the non-unital channel $\Phi$ defined by \eqref{eqphi} is an entanglement-breaking channel iff
\be\label{thm3}
\small
&&\min\{1-\lambda_{i_3},1+\lambda_{i_3}\}\geqslant \nonumber \\
&&\max\{\sqrt{(\lambda_{i_1}+\lambda_{i_2})^2+n_{i_3}^2},
\sqrt{(\lambda_{i_1}-\lambda_{i_2})^2+n_{i_3}^2}\}.
\ee

\textit{Proof}.
Assume $n_1=n_2=0$. The eigenvalues of $\tilde{\rho}$ are given by
\begin{equation}\label{non-unital1}
\begin{array}{rl}
\displaystyle \frac{1}{4}(1-\lambda_3-\sqrt{(\lambda_1-\lambda_2)^2+n_3^2}), \\ [3mm]
\displaystyle \frac{1}{4}(1-\lambda_3+\sqrt{(\lambda_1-\lambda_2)^2+n_3^2}), \\ [3mm]
\displaystyle \frac{1}{4}(1+\lambda_3-\sqrt{(\lambda_1+\lambda_2)^2+n_3^2}), \\ [3mm]
\displaystyle \frac{1}{4}(1+\lambda_3+\sqrt{(\lambda_1+\lambda_2)^2+n_3^2}).
\end{array}
\end{equation}
The eigenvalues of $\tilde{\rho}^{T_2}$ are
\begin{equation}\label{non-unital2}
\begin{array}{rl}
\displaystyle \frac{1}{4}(1-\lambda_3-\sqrt{(\lambda_1+\lambda_2)^2+n_3^2}), \\ [3mm]
\displaystyle \frac{1}{4}(1-\lambda_3+\sqrt{(\lambda_1+\lambda_2)^2+n_3^2}), \\ [3mm]
\displaystyle \frac{1}{4}(1+\lambda_3-\sqrt{(\lambda_1-\lambda_2)^2+n_3^2}), \\ [3mm]
\displaystyle \frac{1}{4}(1+\lambda_3+\sqrt{(\lambda_1-\lambda_2)^2+n_3^2}).
\end{array}
\end{equation}
As $\Phi$ is an EB channel iff the eigenvalues given in \eqref{non-unital1} and \eqref{non-unital2} are nonnegative, we prove the Theorem. For the cases $n_1=n_3=0$ and $n_2=n_3=0$, the proofs are similar.  $\blacksquare$

\section{Relative property of Markovian dynamics}
Due to the importance of  Markovian quantum dynamics in quantum mechanics and in quantum information processing, we analyze relative property of such evolution, in particular, processes of the decoherence, the depolarization, and the homogenization in this section.

The decoherence channels $\Phi_t$ that are modelled as a sequence of collisions of a quantum system with particles of the environment described by a Markovian process can be represented in the form:
\begin{equation*}
\Phi_t=\left(
         \begin{array}{cccc}
           1 & 0 & 0 & 0 \\
           0 & e^{-t/T}\cos \omega t & e^{-t/T}\sin \omega t & 0 \\
           0 & -e^{-t/T}\sin \omega t & e^{-t/T}\cos \omega t & 0 \\
           0 & 0 & 0 & 1 \\
         \end{array}
       \right).
\end{equation*}
In terms of the Markovian semigroup dynamics, the decoherence is induced by the master equation \cite{decoherence,concurrence} $\dot{\rho}=i[H,\rho]+(T/2)[H,[H,\rho]]$. The solution of this equation forms a semigroup of $\Phi_t$. The corresponding singular values $\lambda_i$ in (\ref{eqphi}) related to these channels are given by
\be
\lambda_1(t)&=& e^{-t/T}, \nonumber \\
\lambda_2(t)&=& e^{-t/T}, \\
\lambda_3(t)&=& 1.\nonumber
\ee
Therefore, according to Theorem 1 $\Phi_t$ is not an EB channel, as the inequality (\ref{rhoT2}) is not satisfied.

Another case we consider is the process of a single-qubit depolarization $\Phi_t$ in a specific basis,
which can be represented by the semigroup \cite{quantum}:
\begin{equation*}
\Phi_t=\left(
         \begin{array}{cccc}
           1 & 0 & 0 & 0 \\
           0 & e^{-t/T} & 0 & 0 \\
           0 & 0 & e^{-t/T} & 0 \\
           0 & 0 & 0 & e^{-t/T} \\
         \end{array}
       \right).
\end{equation*}
Using Theorem 1 we have that $\Phi_t$ is an EB channel iff $e^{-t/T}\leqslant \displaystyle{1}/{3}$.

It is obvious that above decoherence and depolarization channels are unital ones.
For an example of nonunital channel we consider the quantum homogenization. The quantum homogenization is a process motivated by the thermodynamical process of thermalization. It describes a system-environment interaction such that the initial state of the system $\rho$ is transformed into the state $\dot{\rho}$ determined by the state of the environment composed of $N$-systems of the same physical origin. The continuous version of the homogenization process can be described by \cite{homo1,homo2}
\begin{equation*}
\small
\Phi_t=\left(
         \begin{array}{cccc}
           1 & 0 & 0 & 0 \\
           0 & e^{-t/T_2}\cos \omega t & e^{-t/T_2}\sin \omega t & 0 \\
           0 & -e^{-t/T_2}\sin \omega t & e^{-t/T_2}\cos \omega t & 0 \\
           w (1-e^{-t/T_1}) & 0 & 0 & e^{-t/T_1} \\
         \end{array}
       \right),
\end{equation*}
\normalsize
where $w$ is the purity of the final state, describing the unitary part of the evolution, $T_1$ is the decay time and $T_2$ is the decoherence time.
This process describes an evolution that transforms the whole Bloch sphere into a single point, i.e., a generalization of an exponential decay. That is, quantum homogenization
is described by a contractive map with the fixed point as the stationary state of the dynamics.
We have the corresponding singular values of $\Phi_t$,
\be
\lambda_1(t)&=& e^{-t/T_2}, \nonumber \\
\lambda_2(t)&=& e^{-t/T_2}, \\
\lambda_3(t)&=& e^{-t/T_1}.\nonumber
\ee
Hence $\Phi_t$ is an EB channel iff
\small
\be \label{H}
(1-w^2)(1-e^{-t/T_1})^2&\geqslant&4e^{-2t/T_2}\nonumber \\
\sqrt{(e^{-t/T_1}+e^{-t/T_2})^2+w^2(1-e^{-t/T_1})^2}&\leqslant&1-e^{-t/T_2}.
\ee
\normalsize

In the homogenization process with the fixed point as a pure state, the homogenization is a nonunital process. In order to analyze the entanglement-breaking property, we denote $f_1=(1-w^2)(1-e^{-t/T_1})^2-4e^{-2t/T_2}$ and $f_2=1-e^{-t/T_2}-\sqrt{(e^{-t/T_1}+e^{-t/T_2})^2+w^2(1-e^{-t/T_1})^2}$.
Then the relation ($\ref{H}$) is equivalent to $f\equiv\min\{f_1, f_2\} \geqslant 0$. It is easy to find that $\Phi_t$ is not always an EB channel and EB capacity of it decreases with the $w$ increasingly (see Fig. 1). In particular, if $w=1$, $\Phi_t$ cannot be EB channel for any $T_1$ and $T_2$. On the other hand, in Fig. 2, we show the influence of decoherence time and the decay time with equal purity of the final state.
\begin{figure}
  \centering
  \includegraphics[width=6.6cm]{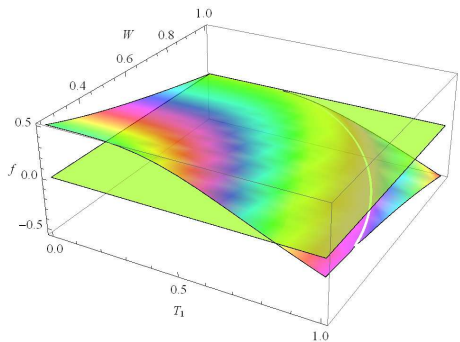}
  \caption{In this figure, we take a special case which the decoherence time is equal to the decay time i.e. $T_1=T_2$. We can see the sloping surface ($f$) through the 0-plane. }
\end{figure}
\begin{figure}
  \centering
  \includegraphics[width=6cm]{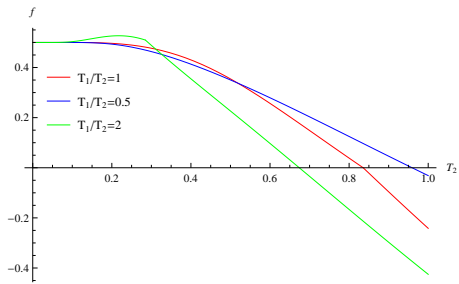}
  \caption{Different curves represent different relations, with $w=0.5$, between $f=\min\{f_1,f_2\}$ and $T_1/T_2$ i.e. the ratio of decoherence time and the decay time. Curves may be under the time axis means the homogenization process may not to be an entanglement-breaking channel.}
\end{figure}

\section{Strong entanglement-breaking channels}
It is natural to ask that if a channel is an EB channel, then how strong it is in breaking the entanglement, and how can one amend the entanglement breaking. For example, one may ask that if
$\Phi \circ \Phi$  is an EB channel, is it possible that
$\Phi \circ \mathcal{F} \circ \Phi $ is not an EB channel
for some suitable unitary operation $\mathcal{F}$ ? Here the role played by $\mathcal{F}$ can be viewed as to amend the EB channel \cite{Amending}.

In order to deal with the above problems, we introduce a concept -- strong entanglement-breaking channels (SEB channels):
the EB channels that can not be amended by local operations.
Although there is a method to amend EB channels via immediately unitary operations \cite{Amending}, we point out that some EB channels can not be amended in this way, i.e., there indeed exist SEB channels .

Assume that $\Phi$ is an EB channel with at least one $\lambda_k=0$ for $k=1,2,3$. Since a quantum channel can be represented as the consecutive applications of a given elementary map $\tilde{\Phi}$ repeated $n$ times, $\Phi=\tilde{\Phi}^n$. We find that the corresponding singular values $\tilde{\lambda}_k$ of $\tilde{M}$ in
 $\tilde{\Phi}\circ\Phi_U\circ\cdots\circ\Phi_U\circ\tilde{\Phi}
 =\left(
  \begin{array}{cc}
   1 & \roarrow{\mathbf{0}} \\
 \tilde{ \roarrow{ \textbf{\textit{n}}}} & \tilde{M} \\
   \end{array}
   \right)$
is also zero, where $\Phi_U$ is an arbitrary unitary channel.
Thus, $\tilde{\Phi}\circ\Phi_U\circ\cdots\circ\Phi_U\circ\tilde{\Phi}$ is always an EB channel.

\noindent $\textbf{Theorem 4.}$
A channel $\Phi$ defined by \eqref{eqphi} with at least one vanishing $\lambda_i$, $i=1,2,3$,
is a strong entanglement-breaking channel.

Although a SEB channel can not be amended by local operations, it may be amended by global operations.
We give an example concerning the SEB channels and their amendment.
Any $d$-dimensional state $\rho$ can be expressed as \cite{RAB}:
\begin{equation*}
\rho=\frac{I}{d}+\frac{1}{2}\sum\limits_{j=0}^{d-2}\sum\limits_{k=j+1}^{d-1}(b_s^{jk}\sigma_s^{jk}+b_a^{jk}\sigma_a^{jk})+
\frac{1}{2}\sum\limits_{l=1}^{d-1}b^{l}\sigma^{l},
\end{equation*}
with $b_{s,a}^{jk}={\rm tr} (\rho\sigma_{s,a}^{jk})$, $b^{l}={\rm tr} (\rho\sigma^{l})$, $\sigma_s^{jk}=|j\rangle\langle k|+|k\rangle\langle j|$,
$\sigma_a^{jk}=-i|j\rangle\langle k|+i|k\rangle\langle j|$ and
$\sigma^{l}=\sqrt{\frac{2}{l(l+1)}}(\sum\limits_{j=0}^{l-1}|j\rangle\langle j|-l|l\rangle\langle l|)$.
Then the state $\rho$ can be rewrite as $\rho=\frac{I}{d}+\roarrow{x}\cdot\roarrow{X}$ with
$\roarrow{x}=(b_s^{01},\ b_a^{01},\ldots,b_s^{d-2,d-1},\ b_a^{d-2,d-1},\ b^1,\ldots, b^{d-1})^{T}$
and $\roarrow{X}=(\sigma_s^{01},\ \sigma_a^{01},\ldots,\sigma_s^{d-2,d-1}, \sigma_a^{d-2,d-1},\ \sigma^1,\ldots,\sigma^{d-1})^{T}$.
Thus, the action of a $d$-dimensional quantum channel $\Phi$ on a qudit state $\rho$ can be expressed by a real $d^2\times d^2$ matrix
$\left(
        \begin{array}{cc}
          p & \roarrow{m}^{T} \\
          \roarrow{n} & M \\
        \end{array}
      \right)$, where $\roarrow{m}, \ \roarrow{n}$ are $(d^2-1)\times 1$ column vector and $M$ is a $(d^2-1)\times(d^2-1)$ matrix.
Then $\Phi=\left(
        \begin{array}{cc}
          1 & \roarrow{0}^{T} \\
          \roarrow{n} & M \\
        \end{array}
      \right)$
maps $\rho=\frac{I}{d}+\roarrow{x}\cdot\roarrow{X}$ to $\Phi(\rho)=\frac{I}{d}+\roarrow{x^{\prime}}\cdot\roarrow{X^{\prime}}$,
where $\roarrow{x^{\prime}}=(b_s^{\prime01},\ b_a^{\prime01},\ldots,b_s^{\prime d-2,d-1},\ b_a^{\prime d-2,d-1},\ b^{\prime1},\ldots,b^{\prime d-1})^{T}$
and $\roarrow{X^{\prime}}=(\sigma_s^{\prime01},\ \sigma_a^{\prime01},\ldots,\sigma_s^{\prime d-2,d-1}, \sigma_a^{\prime d-2,d-1},\ \sigma^{\prime1},\ldots,\sigma^{\prime d-1})^{T}$.
Following the discussion of Ref. \cite{C.King}, we have that up to unitary equivalence, $\Phi$ can be characterized by $2(d-1)$ parameters
as
\begin{equation}\label{stru}
\Phi=\left(
        \begin{array}{ccccc}
          1 & 0 & 0 & \cdots & 0 \\
          n_1 & \lambda_1 & 0 & \cdots & 0 \\
          n_2 & 0 & \lambda_2 & \cdots & 0 \\
          \vdots & \vdots & \vdots & \ddots & \vdots \\
          n_{d^2-1} & 0 & 0 & \cdots & \lambda_{d^2-1} \\
        \end{array}
      \right),
\end{equation}
where $|\lambda_k|\leqslant 1$ are the singular values of $M$.

Now consider a qubit SEB channel $\Phi$ given by
\begin{equation*}
\Phi=\left(\small
                                            \begin{array}{cccc}
                                               1 & 0 & 0 & 0 \\
                                               0 & 0 & 0 & 0 \\
                                               0 & 0 & -\frac{1}{2} & 0 \\
                                               0 & 0 & 0 & \frac{1}{2} \\
                                            \end{array}
                                          \right).
\end{equation*}
Let $\Phi^{\prime}$ be a global operation which is represented by a $16\times16$ matrix defined by \eqref{stru}
with $n_6=n_9=1=\lambda_3=\lambda_5=\lambda_6=\lambda_9=\lambda_{10}=\lambda_{12}=\lambda_{15}=1$ and the other entries are all zero.
Thus,
$$
\Phi^{\prime}\circ(\Phi\otimes I)(|\psi\rangle\langle\psi|)=\left(
        \begin{array}{cccc}
          \frac{1}{2} & 0 & 0 & -\frac{1}{2} \\
          0 & \frac{3}{2} & 1  & 0 \\
          0 & 1 & \frac{3}{2}  & 0 \\
          -\frac{1}{2} & 0 & 0 & \frac{1}{2}.
        \end{array}
      \right)
$$
Therefore, $\Phi$ is amended by a global operation since $\Phi^{\prime}\circ(\Phi\otimes I)(|\psi\rangle\langle\psi|)$ is an entangled state.

From the above example, we see that for SEB channels, there may exist quantum channels $\Phi^{\prime}$ which can increase the entanglement of two-qubit states while acting on them,
i.e., $\Phi^{\prime}\circ(\Phi\otimes T)(|\psi\rangle\langle\psi|)$ is an entangled state.
In other words, a SEB channel may amended by global operations.

\section{Conclusion}
We have investigated both unital and non-unital qubit EB channels and introduced the concept of SEB channels. We have derived both necessary and sufficient conditions for qubit unital channels to be EB channels, as well as for classes of non-unital qubit channels. For a general qubit channel $\Phi$, it has been shown that it cannot be amended by local operations if the rank of the corresponding matrix of $\Phi$ is less than three. For the case that the rank is 4, explicit examples have been presented to illustrate the amendment of the channel. In addition, the EB of Markovian dynamics has been also analyzed.

\section{Acknowledgments}
This work is supported by the NSF of China under Grant No. 11675113 and is supported by the Scientific Research General Program of Beijing Municipal Commission of Education (Grant No.KM 201810011009).

\end{document}